\documentclass[twocolumn,showpacs,preprintnumbers,amsmath,amssymb]{revtex4}
\usepackage{graphicx}% Include figure files
\usepackage{dcolumn}% Align table columns on decimal point
\usepackage{bm}% bold math
\usepackage{epsf}
\usepackage{epsfig}
\usepackage{psfig}
\begin{document}

%\preprint{Heisenberg.limit}

\title{Approaching the Heisenberg limit with two mode squeezed states}

\author{Ole Steuernagel}
 \email{ole@star.herts.ac.uk}
\affiliation{Dept. of Physical Sciences,
University of Hertfordshire, College Lane, Hatfield, AL10 9AB, UK}%

\date{\today} %19.XI.2002}% It is always \today, today,

\begin{abstract}
Two mode squeezed states can be used to achieve Heisenberg limit
scaling in interferometry: a phase shift of $\delta \varphi
\approx 2.76 / \langle N \rangle$ can be resolved. The proposed
scheme relies on balanced homodyne detection and can be
implemented with current technology. The most important
experimental imperfections are studied and their impact
quantified.
\end{abstract}


\maketitle

%\section{\label{sec:level1}First-level heading:\protect\\ The line break was forced \lowercase{via} \textbackslash\textbackslash}
%\subsection{\label{sec:level2}Second-level heading: Formatting}

The best possible phase resolution for an interferometer is given
by the Heisenberg limit for the minimum detectable phase shift
$\delta\varphi = 1/ \langle N\rangle$; here $\langle N\rangle$ is
the average intensity (number of photons or other bosons). Present
optical interferometers typically operate at the shot noise
resolution limit $\delta\varphi\sim 1/\sqrt{\langle N\rangle}$.
Interest in reaching the Heisenberg-limit is great because it
presents a fundamental limit and overcomes the shot-noise limit
leading to potential applications in high resolution distance
measurements, for instance, to detect gravitational
waves~\cite{Scully.buch,{Caves80},Caves81,{Bondurant84},{Yurke86},Xiao87,{Grangier87},Burnett93,{Hillery93},{Jacobson95},{Sanders95},{Ou96},Brif96,Kim98,Dowling98,Gerry02,Soederholm.0204031}.

Known, feasible schemes use degenerate squeezed vacuum combined
with Glauber-coherent light to increase the phase sensitivity
achieving sub-shot noise resolution, but do not reach the
Heisenberg limit~\cite{Xiao87,Grangier87}. Indeed, no practical
scheme has been found that shows {\em scaling like} the Heisenberg
limit $\delta\varphi = \kappa/ \langle N\rangle$ for large
intensities (and preferably a small constant $\kappa$).

More recent publications describing schemes that
theo\-re\-ti\-cally reach the Heisenberg limit have mostly
considered quantum states which are very hard to
synthesize~\cite{Burnett93,{Hillery93},{Jacobson95},{Ou96},Brif96,Dowling98,Gerry02,Soederholm.0204031}
and suggest to use unrealistically high non-linearities to guide
the light through the interferometer~\cite{Jacobson95} or
detectors which have single photon resolution even when dealing
with very many
photons~\cite{Bondurant84,{Yurke86},Burnett93,{Hillery93},{Ou96},Brif96,Kim98,Dowling98,Gerry02,Soederholm.0204031}.

\begin{figure}
\epsfverbosetrue \epsfxsize=3.4in \epsfysize=0.8in
%\epsfverbosetrue \epsfxsize=7.6in \epsfysize=3.5in
%
%
\epsffile{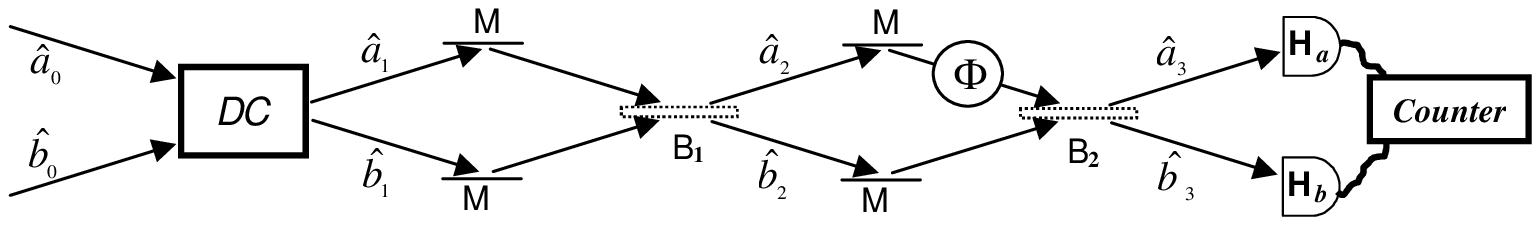}
%\epsffile[076 544 552 730]{Setup.1.eps}
%
\caption{Sketch of the setup: $\sf M$ stands for mirrors and $\sf
B$ for balanced beam-splitters. A non-degenerate parametric
amplifier (down-converter {\it DC}) generates two-mode squeezed
vacuum in modes $\hat a_1$ and $\hat b_1$. These are fed into an
inter\-fero\-meter, with a phase shifter $\Phi$; its output modes
$\hat a_3$ and $\hat b_3$ are detected by homodyne detectors {\sf
H}$_a$ and {\sf H}$_b$ whose signals are multiplied. Note, that
neither the amplifier's pump-laser nor the laser for homodyne
detection is shown, also, typically, the polarization of modes
$\hat a_1$ and $\hat b_1$ does not conform and has to be adjusted.
\label{setup.fig.1}}
\end{figure}

This Letter proposes to use a standard linear two-path
interferometer fed with two mode squeezed vacuum states degenerate
in energy and polarization~\cite{Yurke86,Hillery93}, see
FIG.~\ref{setup.fig.1}. But rather than measuring photon numbers
(intensities) we want to measure the product of the output ports'
quadrature components, i.e. perform balanced homodyne
detection~\cite{Scully.buch,Smithey93}. The only non-linearities
used in the setup proposed here are those of the crystal for
parametric down-conversion to generate the two mode squeezed
vacuum state. It turns out that modest squeezing, i.e. low
intensities, suffice to reach interferometric resolution at
approximately three times the Heisenberg limit
%
%%
%%%
\begin{eqnarray}
\delta \varphi \approx \frac{2.76}{\langle N \rangle} \; .
\label{true.minimal.d.varphi}
\end{eqnarray}
%%%
%%
%
%
%%
%%%
%%%
%%
%

The use of balanced homodyne detection removes the detection
problems mentioned above. Because only well established technology
is required~\cite{Smithey93,Smithey92,Lamas-Linares01} a
proof-of-principle experiment will be immediately possible.

\begin{figure}
\epsfverbosetrue \epsfxsize=3.8in \epsfysize=1.5in
\epsffile[000 100 592 666]{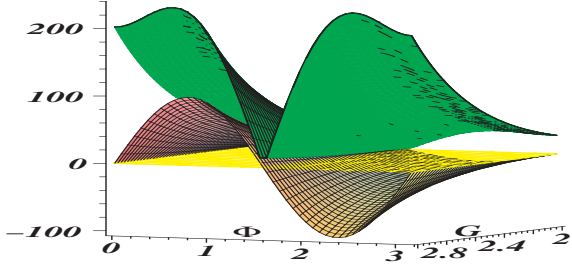}
\caption{\protect{Signal $\langle \hat P \rangle$ and the square
root of the signals second moment $\langle \hat P^2 \rangle$, see
eqns.~(\ref{signal.P}) and~(\ref{second_moment.Pn}), give us an
idea of the behavior of the noise when measuring the homodyne
current product~(\ref{signal.P}).}
\label{fig.2.signal.2ndmoment}}
\end{figure}

In order to derive our main result~(\ref{true.minimal.d.varphi})
we follow the conventions of reference~\cite{Agarwal01}: In the
Heisenberg picture the action of the parametric amplifier is
described by photon operator transformations $\hat a_1= U \hat a_0
+ V \hat b_0^\dag$ and $\hat b_1= U \hat b_0 + V \hat a_0^\dag$
where $U = \cosh G$ and $V = - i \exp(i \xi) \sinh G$ with the
single pass gain $G=g|E_p|L$ and a relative phase $\xi$ which we
will assume to be zero. $L$ is the interaction path length, $ E_p
$ the pump laser's amplitude, and $g$ the gain coefficient
proportional to the nonlinear susceptibility $\chi^{(2)}$ of the
down-conversion medium $ DC$. Beam splitter ${\sf B}_1$ is
described by $\hat a_2 =\exp(i \Phi) (\hat a_1 - i \hat b_1)/\sqrt
2$ and $\hat b_2 =(-i \hat a_1 + \hat b_1)/\sqrt 2$; note that the
interferometric phase shift $\Phi$ in arm $\hat a_2$ is included.
The action of the beam mixer ${\sf B}_2$ is analogously described
by $\hat a_3 =-(\hat a_2 - i \hat b_2)/\sqrt 2$ and $\hat b_3 =(-i
\hat a_2 + \hat b_2)/\sqrt 2$ and the total transformation thus
reads
%
%%
%%%
\begin{eqnarray}
 \hat a_3 &=&\frac{1-e^{i\Phi}}{2}(U \hat a_0 + V \hat b_0^\dag)
 +\frac{i+i e^{i\Phi}}{2} (U \hat b_0 + V \hat a_0^\dag)
\label{trafo.a1.to.a3}
\\
 \hat b_3 &=& \frac{1+e^{i\Phi}}{2i} (U \hat a_0 + V \hat b_0^\dag)
 +\frac{1-e^{i\Phi}}{2} (U \hat b_0 + V \hat a_0^\dag) \; . \;
\label{trafo.a1.to.b3}
\end{eqnarray}
%%%
%%
%
Since we assume that modes $\hat a_0$ and $\hat{b}_0$ are in the
vacuum state, two mode squeezed vacuum in modes $\hat{a}_1$ and
$\hat{b}_1$ results, parameterized by the squeezing or gain
parameter $G$. The corresponding intensity $\langle N \rangle $
is~\cite{Scully.buch}
%
%%
%%%
\begin{eqnarray}
\langle N \rangle \doteq \langle \hat a_3^\dagger \hat a_3 + \hat
b_3^\dagger \hat b_3 \rangle = \langle \hat a_1^\dagger \hat a_1 +
\hat b_1^\dagger \hat b_1 \rangle = 2 \sinh(G)^2  \; .
\label{intensity}
\end{eqnarray}
%%%
%%
%
It is well know that balanced homodyne detection measures the
quadrature components of the monitored fields. We assume a
relative phase of zero between local oscillator and our
interferometric modes $\hat a_3$ and $\hat{b}_3$. In this case the
photo currents of detectors $A$ and $B$ are proportional to the
expectation values of $ \hat a_3^\dagger + \hat a_3 $ and $\hat
b_3^\dagger + \hat b_3$~\cite{Scully.buch,{Gardiner.buch}}. The
product $P$ of the photo-currents is the signal we are interested
in, it amounts to
%
%%
%%%
\begin{eqnarray}
\langle \hat P \rangle = \langle (\hat a_3^\dagger + \hat a_3 )
(\hat b_3^\dagger + \hat b_3 ) \rangle = \sinh(G)\cosh(G)\sin(2
\Phi ). \label{signal.P}
\end{eqnarray}
%%%
%%
%
Note, that we observe a double period in the phase interval
$\Phi=0,...,2\pi$ in Eq.~(\ref{signal.P}) and in
FIG.~\ref{fig.2.signal.2ndmoment} because our signal stems from
the product of two homodyne currents. The corresponding second
moment $\langle \hat P^2 \rangle$ is
%
%%
%%%
\begin{eqnarray}
\langle \hat P^2 \rangle = 1 + [\frac{7}{4} +\cos(2\Phi)
-\frac{3}{4}\cos(4\Phi) ]  \left(\frac{\langle N \rangle^2}{2} +
\langle N \rangle \right), \label{second_moment.Pn}
\end{eqnarray}
%%%
%%
%
where we used the intensity expression~(\ref{intensity}). This
yields the standard deviation $\sigma = \sqrt{\langle \hat
P(\Phi)^2 \rangle - \langle \hat P(\Phi) \rangle^2}$
%
%%
%%%
\begin{eqnarray}
\sigma(\Phi)= \sqrt{ \left( \frac{ \langle N \rangle^2}{2}+\langle
N \rangle \right) [\frac{3}{2} + \cos(2\Phi)
-\frac{\cos(4\Phi)}{2}]+1 }, \label{standard.deviation}
\end{eqnarray}
%%%
%%
%
which is minimal for $\Phi_{min}\doteq \phi =\pi/2$. Consequently
the associated standard expression for the phase resolution limit
$\delta \phi = \sigma(\Phi)/|\partial
P/\partial\Phi|_{\Phi=\pi/2}$ is $1/|\partial
P/\partial\Phi|_{\Phi=\phi} = 1/ \sqrt{\langle N
\rangle^2+2\langle N \rangle} \approx 1/\langle N \rangle$.
%
%%
%%%
%%%
%%
%
This result seems to indicate that we can reach the Heisenberg
limit since the minimal detected phase difference $\delta \phi
\approx 1/ \langle N \rangle$. But an inspection of the behavior
of the second moment of the signal in
FIG.~\ref{fig.2.signal.2ndmoment} shows that the noise varies
greatly in the vicinity of the optimal point $\phi=\pi/2$. We
therefore have to analyze the behavior of the noise-valley around
$\phi$ more closely. It turns out that the rapid growth of noise
away from the optimal point does not let us achieve
Heisenberg-limit resolution but the gradient of the slopes is
sufficiently low to allow for a reduced phase resolution that {\em
scales like} the Heisenberg limit, namely, according to our main
result~(\ref{true.minimal.d.varphi}). Note, that a similar problem
was encountered in reference~\cite{Bondurant84} which was resolved
by the stipulation that the interferometer acted
'phase-conjugated', meaning, when arm $\hat a_2$ lengthens $\hat
b_2$ contracts by the same amount. In the present case this
solution does not help and we have to accept a diminished
performance. To derive our limit~(\ref{true.minimal.d.varphi}),
let us remind ourselves of the standard derivation for the
noise-induced phase-spread that limits interferometric resolution.
%
%%
%%%
\begin{figure}
\epsfverbosetrue \epsfxsize=3.2in \epsfysize=1.6in
%
%
%\epsffile[000 100 592 666]{4.over.N.eps}
\epsffile{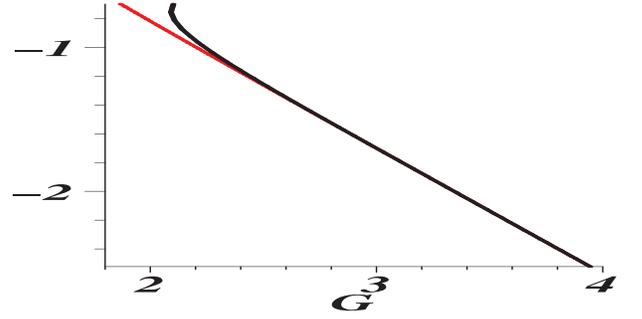}
\caption{Logarithmic plot of minimum phase spread
$\log_{10}(\delta\varphi)$ as a function of the single pass gain
$G$ determined
from~\protect{(\ref{minimal.d.varphi.changing.variance})} (curved
black line) and of four times the inverse intensity
$\log_{10}(4/(2 \sinh(G)^2))$ (straight red line) confirming that
$\delta\varphi=4/(2 \sinh(G)^2)=4/\langle N \rangle$.
\label{4.over.N.fig.3}}
\end{figure}
%%%
%%
%
Assuming that we encounter a noisy signal $ P(\Phi) \pm
\sigma(\Phi)$ with standard deviation $\sigma $ we want to be able
to tell the parameter $\phi$ apart from $\phi + \Delta \phi$. We
therefore require (assuming, for definiteness, that $P\geq 0$ and
growing with increasing $\Phi$) that, according to the
Rayleigh-criterion, $ P(\phi) + \frac{\sigma(\phi)}{2} \lessapprox
P(\phi+ \Delta \phi) - \frac{\sigma(\phi+ \Delta \phi)}{2} $.
Approximating $P(\phi+ \Delta \phi) \approx P(\phi) + \Delta \phi
\cdot \partial P(\phi)/\partial \phi$, assuming equality of left
and right hand side in order to determine the smallest permissible
$\delta \phi$ and that the variance does not change appreciably
$\sigma(\phi+ \Delta \phi) \approx \sigma(\phi)$ this yields the
standard expression for the phase resolution limit $\delta \phi =
\sigma(\phi)/|\partial P/\partial\Phi|_{\Phi=\phi}$. In our case,
however, we need to look at an expression which accounts for the
changing variance; we therefore have to include both variances
$\sigma(\varphi)$ and $\sigma(\varphi+\delta \varphi)$, according
to the above discussion this leads to the modified criterion
%
%%
%%%
\begin{eqnarray}
\delta \varphi = \frac{\sigma(\varphi)+\sigma(\varphi+\delta
\varphi)}{2} \cdot \frac{1}{|\partial P/\partial
\Phi|_{\Phi=\varphi} }\; .
\label{minimal.d.varphi.changing.variance}
\end{eqnarray}
%%%
%%
%
Choosing the optimal working point $\varphi=\pi/2$, this yields an
implicit equation for $\delta \varphi$ which is not too easy to
solve in the general case but for sufficiently high intensities
($G>2.5 \hookrightarrow \langle N \rangle > 2 \sinh(2.5)^2 \approx
73 $ photons) we find $\delta \varphi = 4/ \langle N \rangle$.
This is illustrated by FIG.~\ref{4.over.N.fig.3} and can be
verified by direct substitution
into~(\ref{minimal.d.varphi.changing.variance}). Because in our
scheme the noise is phase sensitive it only works at particular
phase settings (odd multiples of $\pi/2$, see
FIG.~\ref{fig.2.signal.2ndmoment}) and our setup has to include a
feedback mechanism -- not mentioned in FIG.~\ref{setup.fig.1}.

{\em Robustness and further increase in sensitivity:}\\
Having shown that our scheme allows for Heisenberg-limit--like
scaling in interferometric sensitivity we would also like to look
at its sensitivity to experimental imperfections. Balanced
homodyne detection amplifies quantum features to the classical
level~\cite{Scully.buch}. For strong fields detector losses can be
kept small~\cite{{Smithey93},{Smithey92},Gardiner.buch} and will
therefore not be discussed further.

More importantly, losses and imbalances in the state preparation
and interferometric part of the setup sketched in
FIG.~\ref{setup.fig.1} deserve consideration. The main question we
want to address is whether the introduction of experimental
imperfections leads to a gradual loss of performance or whether we
might be unlucky and a qualitative change in behavior results from
any minute imperfection. It turns out that the former is the case,
yet, experimental demands on the state preparation part of the
setup are very high.
%
%%
%%%
\begin{figure}
\centering
\includegraphics[width=1.5in,height=1in]{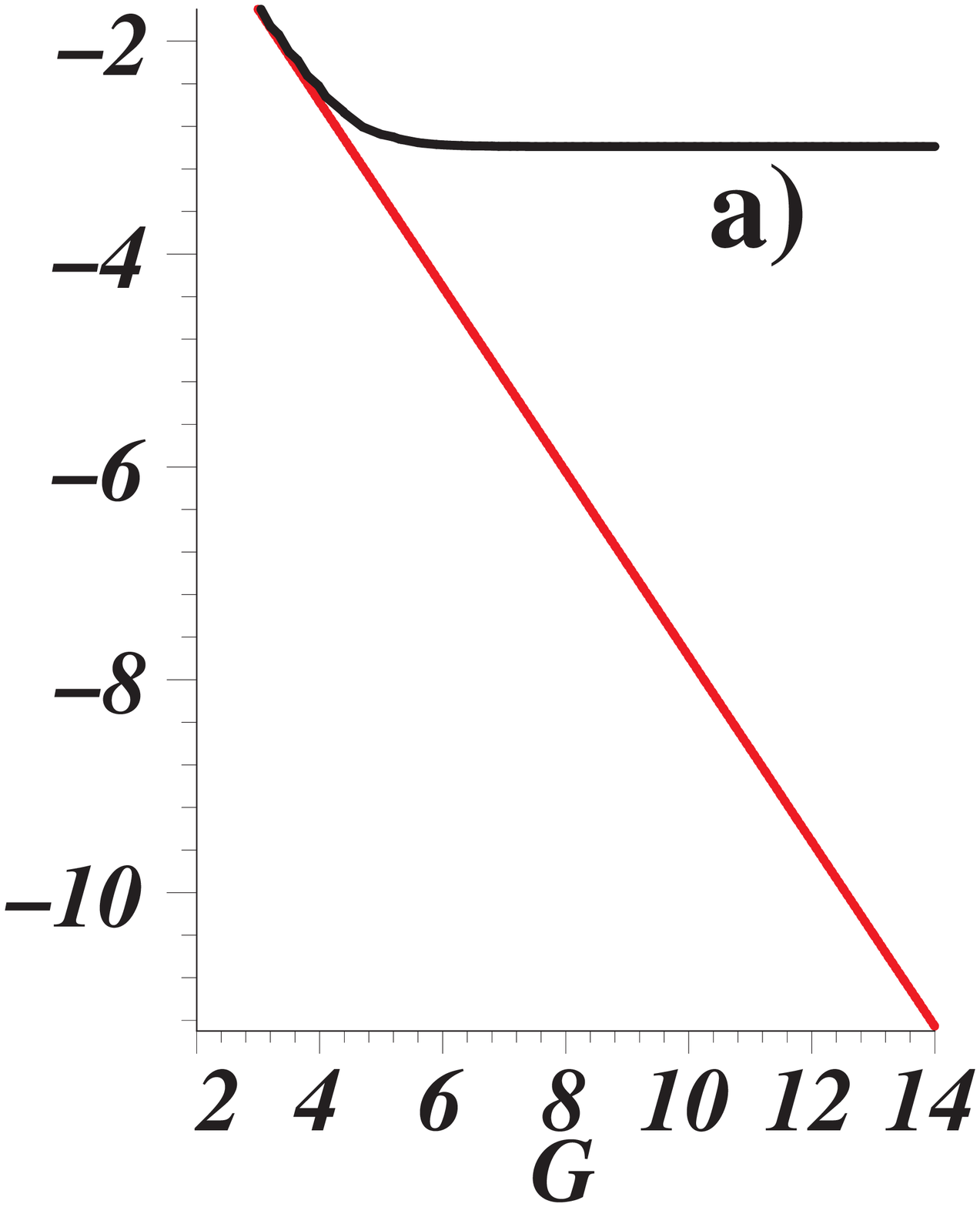}
\hspace{0.1in}%
\includegraphics[width=1.5in,height=1in]{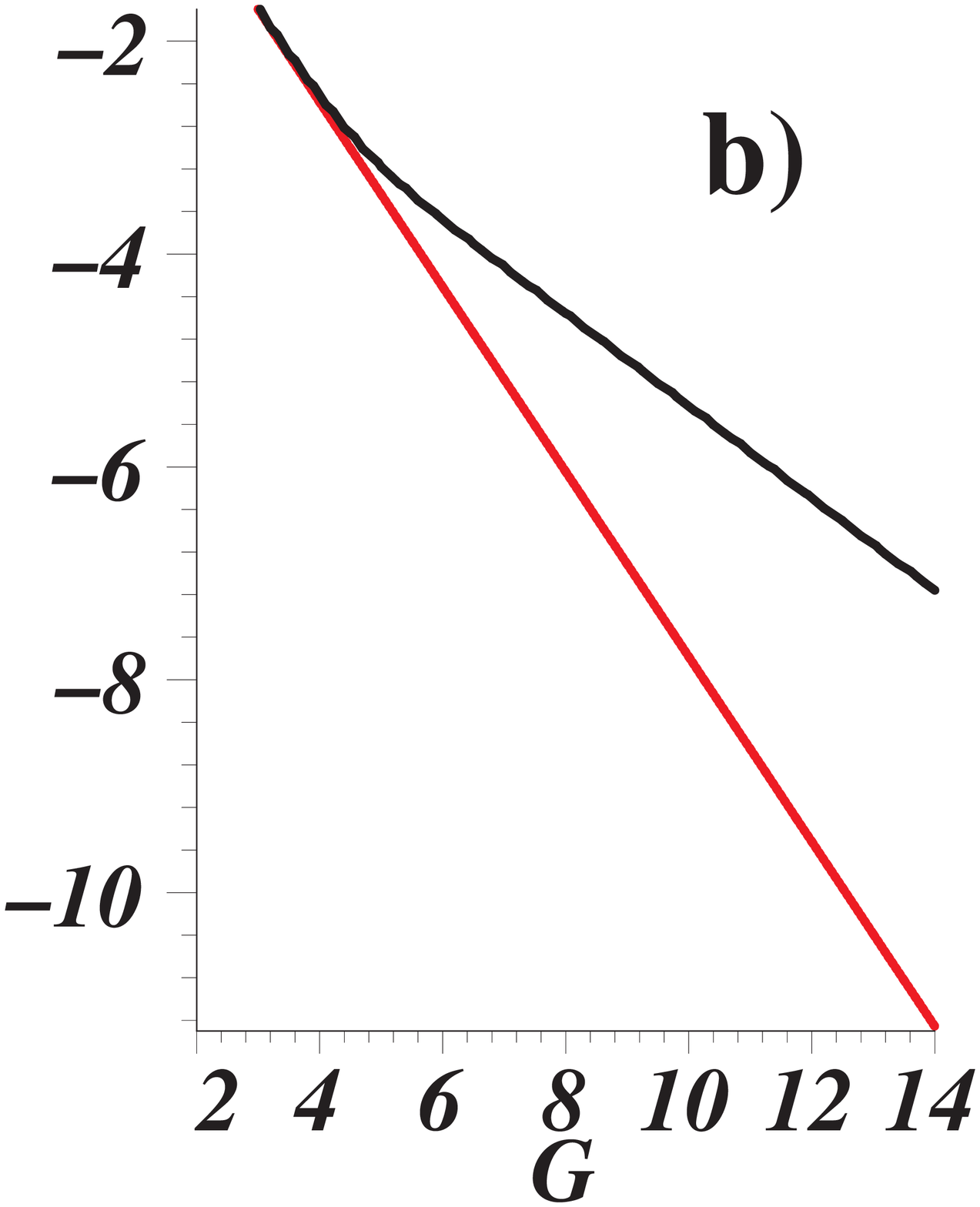}
\\
\includegraphics[width=1.5in,height=1in]{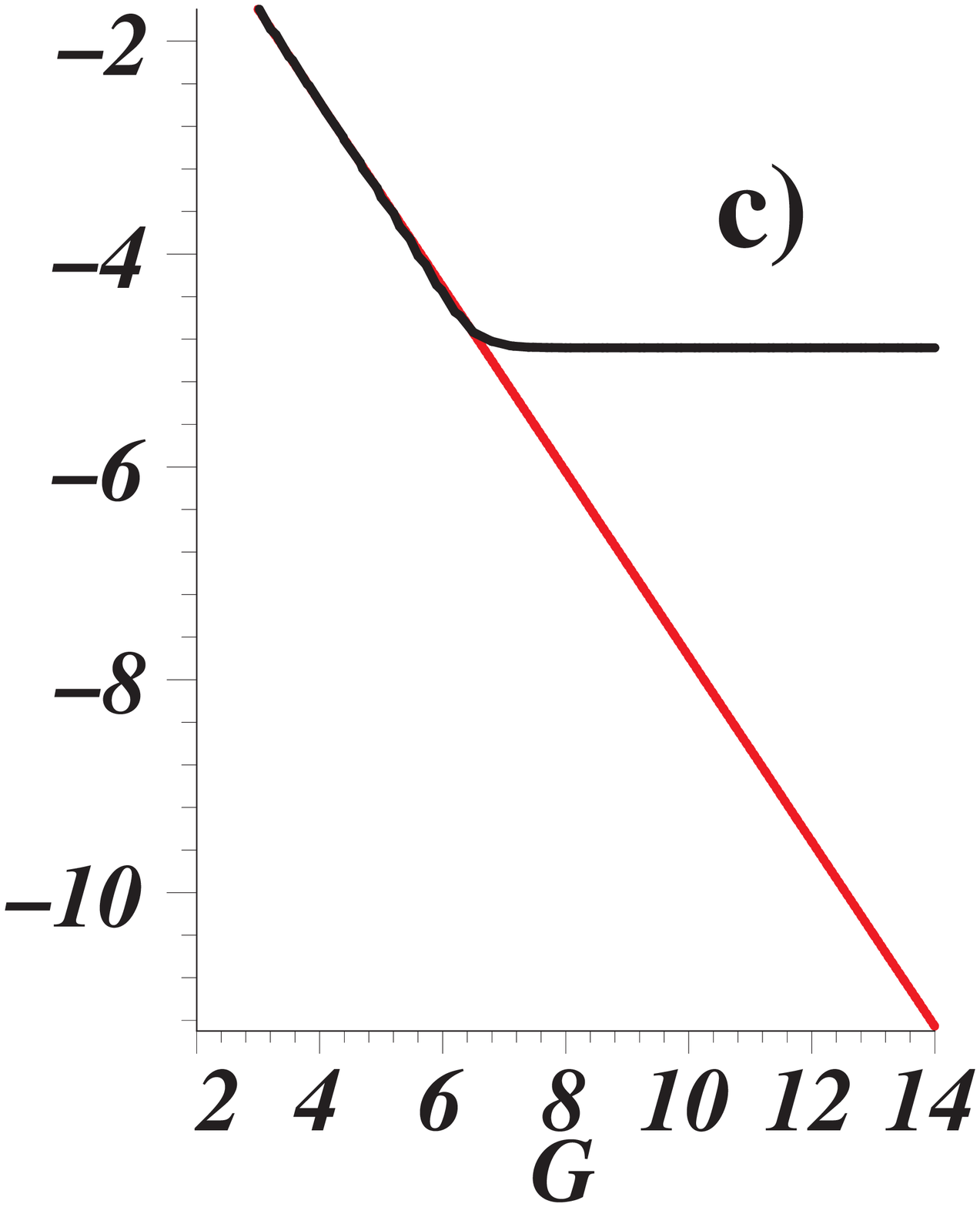}
\hspace{0.1in}%
\includegraphics[width=1.5in,height=1in]{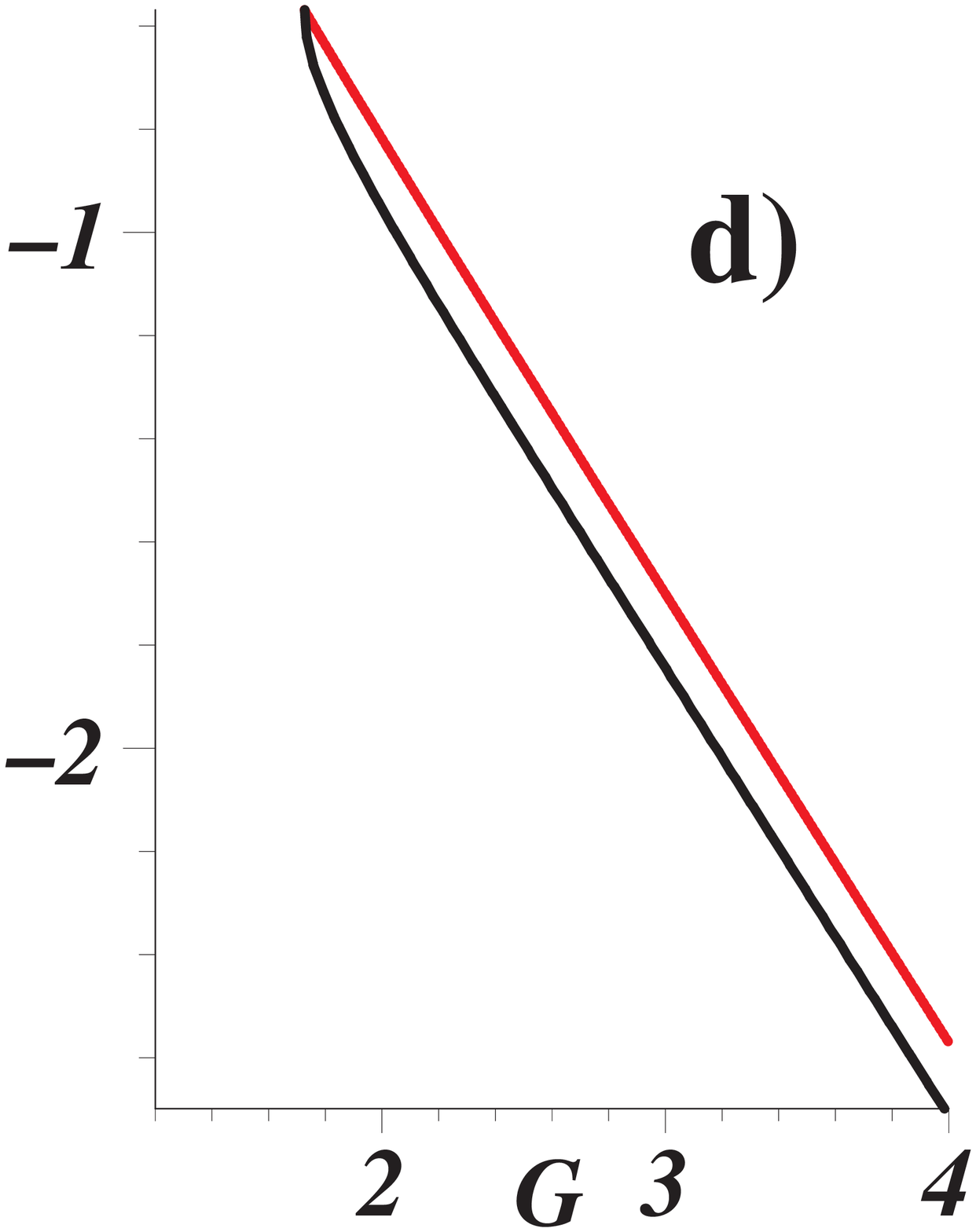}
%
%\epsfverbosetrue \epsfxsize=3.2in \epsfysize=1.6in
%\epsffile{4.over.N.eps}
%
%
\caption{Logarithmic plot of minimum phase spread
$\log_{10}(\delta\varphi)$ determined
from~\protect{(\ref{minimal.d.varphi.changing.variance})} (curved
black lines) and of four times the inverse intensity
$\log_{10}(4/(2 \sinh(G)^2))$ (straight red lines) comparing
various loss mechanisms: a) $\alpha_1 = \beta_1 =\pi/300$,  b)
$\alpha_2 = \beta_2 =\pi/300$, c) $\Delta_1 = 1/300000$, d)
$\Delta_2 =-1/3$. Note the different orders of magnitude in the
values of $\Delta_1$ and $\Delta_2$ and the fact that 'd)',
surprisingly, shows a slight {\em improvement} in performance for
a large negative imbalance $\Delta_2=-1/3$.
\label{4.over.N.figs.4}}
\end{figure}
%%%
%%
%

FIG.~\ref{4.over.N.figs.4} compares the various cases for losses
and imbalances and shows that the system is more forgiving for
losses in the interferometer part than in the state preparation
part: the utilized quantum state has to be prepared with great
skill but the scheme is comparatively robust to imperfections of
the inter\-fero\-meter. When all imperfections are studied
simultaneously their effects add up, i.e., tend to be dominated by
the largest effect(s).

Let us first consider losses in the state preparation part of the
setup, i.e. losses in modes $\hat a_1$ and $\hat b_1$ extending
from inside the crystal to the first beam-splitter $ {\sf B}_1$.
They are described by the mode-transformations $\hat a_1 \mapsto
\cos(\alpha_1) \hat a_1 + \sin(\alpha_1) \hat u_1$ and $\hat b_1
\mapsto \cos(\beta_1) \hat b_1 + \sin(\beta_1) \hat v_1 $ plus
subsequent tracing over the loss modes (not mentioned) and the
admixed vacuum modes $\hat u_1$ and $\hat v_1$. It turns out that
the qualitative picture does not depend much on the details such
as whether the loss parameters $\alpha_1$ and $\beta_1$ are equal
or the losses occur in one channel only. Thus, with experiments in
mind, let us assume symmetric losses, namely $\alpha_1 = \beta_1 =
\pi/300 \approx 0.01$ leading to $0.0001=0.01$~\% losses in both
channels. For example, mode-mismatch at the beam splitter $ {\sf
B}_1$ leads to such symmetrical admixture of vacuum. This scheme
is very sensitive to losses in the state preparation part of the
setup and shows saturation of performance, see
FIG.~\ref{4.over.N.figs.4}~a): to gain an order of magnitude in
performance $\alpha_1$ and $\beta_1$ have to be decreased by half
an order of magnitude, namely, $\delta\varphi$ saturates at about
$9 \alpha_1^2$.

Losses in the interferometric part of the setup (modes $\hat a_2$
and $\hat b_2$) are analo\-gous\-ly described by loss parameters
$\alpha_2$ and $\beta_2$ which parameterize the admixture of two
more vacuum modes $\hat u_2$ and $\hat v_2$ to the path modes
$\hat a_2$ and $\hat b_2$. FIG.~\ref{4.over.N.figs.4}~b)
illustrates the greater tolerance of our scheme to losses in the
interferometer part of the setup. For the same loss values,
$\alpha_2 = \beta_2 = \pi/300$ as for $\alpha_1$ and $\beta_1$
before, the scheme shows a relatively better performance, indeed,
the performance does not show saturation at all. Instead, at the
threshold intensity  $ \approx 4/(9 \alpha^2)$ it switches from
$\delta \varphi \approx 4/\langle N \rangle $ to the poorer
scaling $\delta \varphi = \kappa / \sqrt{\langle N \rangle} $,
maintaining the ground it has gained. Namely, beyond the threshold
intensity we find $\delta \varphi \approx 6 \, |\alpha_2| /
\sqrt{\langle N \rangle} $.

Analogously to the case of losses, the scheme is also much more
sensitive to imbalances of the first beam-splitter ${\sf B}_1$
than to those of ${\sf B}_2$, which is described by
%
%%
%%%
\begin{eqnarray}
%
%\left[\begin {array}{cc} \hat a_3 \\ \hat b_3 \end {array}\right]=
\hat{\sf B}_2 \left[\begin {array}{cc} \hat a_2 \\ \hat b_2 \end
{array}\right]
 = \left[\begin {array}{lr} -\cos(\frac{\pi}{4} + \Delta_2) & i
\sin(\frac{\pi}{4} + \Delta_2)\\\noalign{\medskip}-i
\sin(\frac{\pi}{4} + \Delta_2) & \cos(\frac{\pi}{4} + \Delta_2)
\end {array}\right]
\left[\begin {array}{cc} \hat a_2 \\ \hat b_2 \end {array}\right]
\label{imbalance.Delta2}
\end{eqnarray}
%%%
%%
%
conforming with the case $\Delta_2 = 0$ used in the derivation of
Eqs.~(\ref{trafo.a1.to.a3}) and~(\ref{trafo.a1.to.b3}). The
imbalance in transformation $\hat {\sf B}_1$ is described, in full
analogy, by an imbalance angle $\Delta_1$. Similarly to the case
of $\alpha_1$ a non-zero $\Delta_1$ leads to saturation: for
positive imbalances the saturation level is $\delta\varphi \approx
4 \Delta_1 $ and it is $\delta\varphi \approx 12 |\Delta_1| $ for
negative $\Delta_1 $.

It turns out that variation of $\Delta_2$ modifies the coefficient
$\kappa$ but not the scaling exponent in $\delta \varphi = \kappa
/\langle N \rangle $; as mentioned above, for $\Delta_2 = 0$ we
find $\delta \varphi = 4 /\langle N \rangle $. Note that,
surprisingly, a large (negative) imbalance~$\Delta_2$, as is
displayed in FIG.~\ref{4.over.N.figs.4}~d), yields a small {\em
increase} in performance quality ($\kappa$ is being reduced). I
cannot explain this finding and I think it deserves further
investigation and might even lead to a trick to reduce the scaling
reported here down to the Heisenberg-limit
$\delta\varphi=1/\langle N\rangle$. Variation of the value of the
imbalance parameter $\Delta_2$ {\em alone}, leads to an optimal
value for the imbalance of approximately $\Delta_2 \approx -0.2375
\hat = - 13.61^o$ and to our central result
Eq.~(\ref{true.minimal.d.varphi}). This probably is the best our
scheme can offer.

Over recent years a consensus has emerged that a sharp photon
number distribution is needed to reach the
Heisenberg-limit~\cite{{Sanders95},{Dowling98},{Gerry02},{Soederholm.0204031}}.
It was therefore even concluded that the perceived need of a
sub-poissonian photon number distribution renders the two-mode
squeezed vacuum state unsuitable for interferometry because of its
super-poissonian thermal photon number
distribution~\cite{{Sanders95},{Gerry02}}; in the light of these
claims our central result Eq.~(\ref{true.minimal.d.varphi}) is
rather surprising.

Note, that we did not discuss a criterion for the power of the
pump beam driving the parametric down-conversion source and of the
power needed for the strong local oscillator fields necessary to
perform the balanced homodyne measurements {\sf H}$_a$ and {\sf
H}$_b$, see FIG.~\ref{setup.fig.1}. If this is included, the
effective performance of our scheme could be reclassified as less
efficient, yet, it remains a scheme with Heisenberg-limit--like
scaling. The penalty to pay is not too large for large intensities
because the local oscillator's shot-noise-to-signal-ratio
diminishes with increasing signal strength thus yielding very
accurate homodyning signals. In this context I would also like to
mention that there are promising recent ideas for efficient and
bright down-conversion sources~\cite{DeRossi02}.

Also note, that the considerations of this paper might turn out to
be of importance for atom-beam interferometry~\cite{Dowling98}
since four-wave mixing has been reported yielding correlated atom
beams in states similar to the two-mode squeezed vacuum states
discussed here~\cite{Vogels02}.

{\em Conclusions:} We have found that bosonic two-mode squeezed
states can be used in an interferometer to achieve phase
resolution near the Heisenberg-limit, see
Eq.~(\ref{true.minimal.d.varphi}). This only works at particular
phase settings, the noise is phase sensitive and the setup
therefore needs a feedback mechanism. The degrading influence of
experimental imperfections is analyzed and it is shown that
requirements on the state-preparation part of the setup are very
stringent. On the other hand our scheme is more robust with
respect to imperfections of the interferometer part of the setup
and it does not suffer from single-photon detection problems
because it relies on balanced homodyne-detection.

\begin{acknowledgments}
I wish to thank Janne Ruostekoski for reading of this manuscript.
\end{acknowledgments}

\end{document}